\documentclass[%
preprint,
 amsmath,amssymb,
 aps,
]{revtex4-1}

\usepackage{graphicx}
\usepackage{dcolumn}
\usepackage{bm}
\usepackage{hyperref}

\begin{document}

\title{Non-maximal $\theta_{23}$, large $\theta_{13}$ and tri-bimaximal $\theta_{12}$ via \\quark-lepton complementarity at next-to-leading order}%

\author{Junpei Harada}
 \email{harada@sci.niihama-nct.ac.jp}
\affiliation{%
 Niihama National College of Technology, Niihama 792-8580, Japan
}%

\date{July 24, 2013}

\begin{abstract}
We present analytical formulae for the neutrino mixing angles at the next-to-leading order in the quark-lepton complementarity, and show that higher order corrections are important to explain the observed pattern of neutrino mixing. In particular, the next-to-leading order corrections 1) lead to a deviation of $\theta_{23}$ from maximal mixing, 2) reduce the predicted value of $\sin^2 2\theta_{13}$ by $9.8\%$, 3) provide the same value of $\sin^2 \theta_{12}$ as that of the tri-bimaximal mixing.
\end{abstract}


\maketitle
\section{Introduction and motivation \label{sec:intro}}
The main recent developments on neutrino mixing\cite{Altarelli:2004za,Mohapatra:2005wg,Mohapatra:2006gs,Strumia:2006db,GonzalezGarcia:2007ib} are related to the relatively large value of $\theta_{13}$, which was measured by recent experiments\cite{Abe:2011sj,Adamson:2011qu,Abe:2012tg,An:2012eh,Ahn:2012nd},  and indications of significant deviation of $\theta_{23}$ from maximal mixing\cite{Altarelli:2012vq,Smirnov:2012ei}. In particular, non-maximal $\theta_{23}$ is strongly indicated by recent data, but the sign of $\theta_{23} - \pi/4$ is not yet determined. Results of global analysis are given in refs.\cite{Tortola:2012te,Fogli:2012ua,GonzalezGarcia:2012sz}.

To explain the observed pattern of neutrino mixing, the quark-lepton complementarity (QLC) has been widely investigated in the literature\cite{Raidal:2004iw,Minakata:2004xt,Frampton:2004vw,Kang:2005as,Xing:2005ur,Datta:2005ci,Antusch:2005ca,Minakata:2005rf,Jarlskog:2005jn,Everett:2005ku,Harada:2005km,Dighe:2006zk,Chauhan:2006im,Hochmuth:2006xn,Schmidt:2006rb,Plentinger:2006nb,Plentinger:2007px,Frampton:2008ep,Xing:2009eg,Zheng:2010kp,Barranco:2010we,Patel:2010hr,Shimizu:2010pg,Ahn:2011yj,Li:2011ag,Zhang:2012zh,Zhang:2012pv}. In particular, much attention has been paid to the class of models, $U_{\mbox{\scriptsize PMNS}} = V_{\mbox{\scriptsize CKM}}^\dagger V_M$, which can be obtained in grand unified theories (GUTs). The correlation matrix $V_M$ is simply defined by the product of the CKM and PMNS mixing matrices. In general $V_M$ is not determined by theories, because there is no relation between the Dirac and the Majorana mass operators~\cite{Harada:2005km}. In this class of models, the observed two large mixing angles $\theta_{12}$ and $\theta_{23}$ indicate that $V_M$ has two large mixing angles because all the mixing angles in $V_{\mbox{\scriptsize CKM}}$ are small. As the simplest possibility, for example, we can take $V_M$ being the bimaximal mixing matrix $V_{bm}$,
\begin{align}
	U_{\mbox{\scriptsize PMNS}} = V_{\mbox{\scriptsize CKM}}^\dagger V_{bm}.
	\label{eq:PMNS-CKM}
\end{align}

In this paper we only consider this minimal model. In ref.\cite{Harada:2005km}, expanding $V_{\mbox{\scriptsize CKM}}$ to ${\cal O}(\lambda^4)$ (where $\lambda \equiv \sin \theta_{\mbox{\scriptsize C}} \approx 0.2253$), it is found that
\begin{align}
	\sin^2 2\theta_{13} &= 2\lambda^2 + {\cal O}(\lambda^4) = 0.102 + {\cal O}(\lambda^4), \label{eq:13mixingLO} \\
	\sin^2 2\theta_{23} &= 1 + {\cal O}(\lambda^4), \label{eq:23mixingLO} \\
	\sin^2 2\theta_{12} &= 1 - 2\lambda^2 + {\cal O}(\lambda^4) = 0.898 + {\cal O}(\lambda^4). \label{eq:12mixingLO}
\end{align}
These correspond to $\theta_{13} \simeq \theta_{\mbox{\scriptsize C}}/\sqrt{2} \simeq 9^\circ$, $\theta_{23} \simeq 45^\circ$ and $\theta_{12} \simeq 36^\circ$, which are consistent with experimental results at leading order approximation. In particular, it is interesting that the predicted value of 1-3 mixing, $\theta_{13} \simeq 9^\circ$,  has been confirmed by recent experiments\cite{Abe:2011sj,Adamson:2011qu,Abe:2012tg,An:2012eh,Ahn:2012nd}.

However, the next-to-leading order corrections, the ${\cal O}(\lambda^4)$ terms in eqs.\eqref{eq:13mixingLO}--\eqref{eq:12mixingLO}, should be calculated, because:
\begin{itemize}
	\item the ${\cal O}(\lambda^4)$ corrections may not be enough small with respect to the recent experimental errors. 
		Particularly, the Daya Bay Collaboration reported the precise value at $1\sigma$\cite{An:2012eh},
		\begin{align}
			\sin^2 2\theta_{13} = 0.092 \pm 0.016(\mbox{stat}) \pm 0.005(\mbox{syst}) 
			\label{eq:Daya-Bay}
		\end{align}	
		where the magnitude of systematic errors is the same order of that of $\lambda^4$; 
	\item the same value of 1-3 mixing, $\theta_{13} \simeq 9^\circ$,  can be obtained in various models with different schemes: flavor symmetries, texture, ans\"{a}tz etc\cite{Smirnov:2012ei}. 
		To distinguish models, we need a more precise prediction;
	\item a deviation of $\theta_{23}$ from maximal mixing is the ${\cal O}(\lambda^4)$ correction in eq.\eqref{eq:23mixingLO}. 
		To determine the magnitude of deviation, the ${\cal O}(\lambda^4)$ corrections should be calculated;
	\item in the earlier works, eq.\eqref{eq:PMNS-CKM} was analyzed numerically. Analytical formulae of higher-order terms are usually neglected; therefore, it is very unclear that which parameter is relevant at each order. 
\end{itemize}

Motivated by these points, in this paper we perform analytical calculations of the ${\cal O}(\lambda^4)$ corrections. We show that the next-to-leading order corrections
\begin{itemize}
	\item reduce the value of $\sin^2 2\theta_{13}$ from 0.102 to 0.092, which is completely consistent with the result of Daya Bay Collaboration of eq.\eqref{eq:Daya-Bay}. 
	\item lead to a deviation of $\theta_{23}$ from maximal mixing. The predicted value is $\sin^2 \theta_{23} = 0.446 + {\cal O}(\lambda^6)$, which corresponds to $\theta_{23} \simeq 41.9^\circ$. 
	\item provide the value of $\sin^2 \theta_{12}$ of 0.335, which is very close to the predicted value of tri-bimaximal (TBM) mixing of 0.333. 
\end{itemize}

\section{Neutrino mixing angles \label{sec:Mixingangles}}
\subsection{1-3 mixing $\theta_{13}$ \label{subsec:theta_13}}
The recent experimental results on $\theta_{13}$ from two accelerator experiments, T2K\cite{Abe:2011sj} and MINOS\cite{Adamson:2011qu}, and from three reactor experiments, Double-Chooz\cite{Abe:2012tg}, Daya Bay\cite{An:2012eh} and RENO\cite{Ahn:2012nd} were very important developments in neutrino physics. 
The global fit value of $\theta_{13}$ is rather large, $\theta_{13} \simeq 9^\circ$ for the normal and inverted mass hierarchy, and $\theta_{13} = 0$ is now excluded at more than 10$\sigma$.

The 1-3 mixing angle $\theta_{13}$ is given by
\begin{align}
	\sin^2 2\theta_{13} = 4 |U_{e3}|^2\left( 1 - |U_{e3}|^2\right). \label{eq:sin22theta_13}
\end{align}
Eq.\eqref{eq:sin22theta_13} indicates that $\theta_{13}$ is determined only by $U_{e3}$. 
In eq.\eqref{eq:sin22theta_13}, $\theta_{13}$ is a parameter defined by the standard parametrization, and the $U_{e3}$ is the matrix element of any parametrization of $U_{\mbox{\scriptsize PMNS}}$\cite{Ohlsson:2002rb}. 

Using eq.\eqref{eq:Ue3^2} in the appendix, we find that 
\begin{align}
	\sin^2 2\theta_{13} 
	&= 2\lambda^2 - \left\{ 1 + 4A (1 - \overline{\rho})\right\} \lambda^4 + {\cal O}(\lambda^6), \nonumber \\
	&= \underbrace{0.1016}_{\lambda^2 \ \mbox{\scriptsize term}} - \underbrace{0.0098}_{\lambda^4 \ \mbox{\scriptsize term}} + {\cal O}(\lambda^6), \nonumber \\
	&= 0.092 + {\cal O}(\lambda^6), \label{eq:13mixingNLO}
\end{align}
where we have used the best-fit values of $\lambda$, $A$ and $\overline{\rho}$ of eq.\eqref{eq:Wolfenstein_values}. It was confirmed that varying values in the error range in eq.\eqref{eq:Wolfenstein_values} are almost negligible. Eq.\eqref{eq:13mixingNLO} corresponds to $\theta_{13} \simeq 8.8^\circ$ (an another solution of $\theta_{13} \simeq 81.2^\circ$ is excluded, because $\sin^2 \theta_{13} = |U_{e3}|^2 < 0.5$).

The key points of eq.\eqref{eq:13mixingNLO} are:
\begin{itemize}
	\item  $\sin^2 2\theta_{13}$ is determined only by one parameter $\lambda$ at leading order, and three parameters $\lambda$, $A$ and $\overline{\rho}$ are relevant at next-to-leading order. The  $\overline{\eta}$ is irrelevant up to ${\cal O}(\lambda^6)$;
	\item the next-to-leading order corrections reduce the value of $\sin^2 2\theta_{13}$ from 0.102 to 0.092, which is completely consistent with the result of Daya Bay Collaboration of eq.\eqref{eq:Daya-Bay}. Thus, the value of $\sin^2 2\theta_{13}$ becomes smaller by $9.8 \%$ due to ${\cal O}(\lambda^4)$ correction terms;
	\item the magnitude of the ${\cal O}(\lambda^4)$ correction terms is approximately twice larger than that of the systematic errors of Daya Bay Collaboration, and therefore the ${\cal O}(\lambda^4)$ corrections cannot be negligible.
\end{itemize}

\begin{figure}[tbp]
\centering
\hfill
\includegraphics[width=1\textwidth,origin=c]{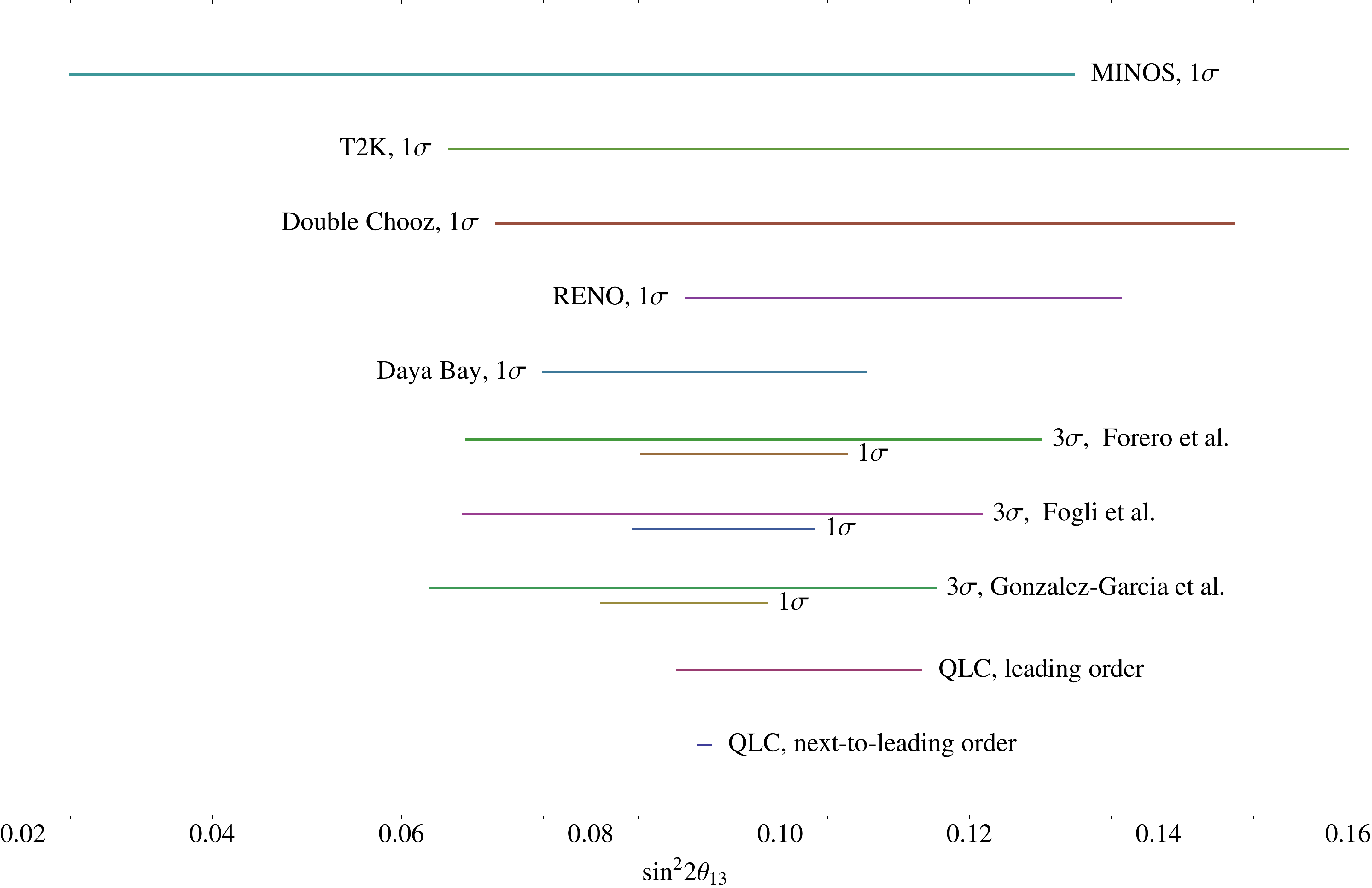}
\caption{\label{fig:13mixing} Determination of the 1-3 mixing. Shown are the results from T2K\cite{Abe:2011sj}, MINOS\cite{Adamson:2011qu}, Double Chooz\cite{Abe:2012tg}, Daya Bay\cite{An:2012eh}, RENO\cite{Ahn:2012nd} and global fits of Forero et al.\cite{Tortola:2012te}, Fogli et al.\cite{Fogli:2012ua} and Gonzalez-Garcia et al.\cite{GonzalezGarcia:2012sz} for the normal hierarchy case (in the inverse hierarchy case the values do not differ by much). The QLC predictions are shown at leading and next-to-leading order.}
\end{figure}

In figure~\ref{fig:13mixing}, we summarize the theoretical and experimental values of $\sin^2 2\theta_{13}$. 
The results from T2K\cite{Abe:2011sj} and MINOS\cite{Adamson:2011qu} are shown for $\sin^2 \theta_{23}=0.4$ and $\delta=\pi$. For the QLC predictions, eq.\eqref{eq:13mixingLO} is shown for $\sin^2 2\theta_{13} = 0.102 \pm 5 \lambda^4$ (leading order), and eq.\eqref{eq:13mixingNLO} is shown for $\sin^2 2\theta_{13} = 0.092 \pm 5\lambda^6$ (next-to-leading order), which correspond to $0.089 < \sin^2 2\theta_{13} < 0.114$ and $0.091 < \sin^2 2\theta_{13} < 0.093$, respectively. Figure~\ref{fig:13mixing} shows that the QLC prediction at next-to-leading order is consistent with the results of five experiments\cite{Abe:2011sj,Adamson:2011qu,Abe:2012tg,An:2012eh,Ahn:2012nd} and three global fits\cite{Tortola:2012te,Fogli:2012ua,GonzalezGarcia:2012sz} at the $1\sigma$ level. 

\subsection{2-3 mixing $\theta_{23}$ \label{subsec:theta_23}}

The recent global analysis indicates that there is a solid deviation of $\theta_{23}$ from maximal mixing\cite{Altarelli:2012vq,Smirnov:2012ei}. At present, the sign of $\theta_{23} - \pi/4$ is not yet determined\cite{Tortola:2012te, Fogli:2012ua, GonzalezGarcia:2012sz}. 

Eq.\eqref{eq:23mixingLO} shows that the 2-3 mixing is maximal at leading order. 
Eq.\eqref{eq:23mixingLO} also shows that a deviation from maximal mixing is not ${\cal O}(\lambda^2)$ but the ${\cal O}(\lambda^4)$ effect. Therefore, a deviation is not so large. 

The 2-3 mixing angle $\theta_{23}$ is given by
\begin{align}
	\sin^2 2 \theta_{23} = \frac{4 |U_{\mu 3}|^2 |U_{\tau 3}|^2}{\left( 1 - |U_{e3}|^2\right)^2}. \label{eq:sin22theta_23}
\end{align}
Eq.\eqref{eq:sin22theta_23} shows that $\theta_{23}$ is determined by three matrix elements, $U_{\mu 3}$, $U_{\tau 3}$ and $U_{e3}$. Of these elements, two are independent because of the unitarity condition $|U_{e3}|^2 + |U_{\mu 3}|^2+|U_{\tau 3}|^2=1$.

From eqs.\eqref{eq:Umu3^2}, \eqref{eq:Utau3^2} and \eqref{eq:1/(1-Ue32)^2}, we find that 
\begin{align}
	\sin^2 2\theta_{23} &
	= 1 - 4 \left(\frac{1}{4} +  A \right)^2 \lambda^4  + {\cal O}(\lambda^6), \nonumber \\
	&= 1 - \underbrace{0.0116}_{\lambda^4 \ \mbox{\scriptsize term}} + {\cal O}(\lambda^6), \nonumber \\
	&= 0.988 + {\cal O}(\lambda^6). \label{eq:23mixingNLO}
\end{align}
Eq.\eqref{eq:23mixingNLO} indicates that the value of $\sin^2 2\theta_{23}$ is determined only by two parameters, $\lambda$ and $A$ at next-to-leading order. The $\overline{\rho}$ and $\overline{\eta}$ are irrelevant up to ${\cal O}(\lambda^6)$. 

To determine the sign of $\theta_{23} - \pi /4$, we calculate $\sin^2 \theta_{23} = |U_{\mu 3}|^2/(1-|U_{e3}|^2)$ .
From \eqref{eq:1/1-Ue32}, we find that 
\begin{align}
	\sin^2 \theta_{23} 
	&= \frac{1}{2} - \left(\frac{1}{4} +  A \right)\lambda^2   
		- \frac{1}{2}\left(\frac{1}{4} + A\overline{\rho}\right)\lambda^4
		+{\cal O}(\lambda^6), \nonumber \\
	&= \frac{1}{2} - \underbrace{0.0539}_{\lambda^2 \ \mbox{\scriptsize term}} 
		     - \underbrace{0.0005}_{\lambda^4 \ \mbox{\scriptsize term}} 
		+ {\cal O}(\lambda^6), \nonumber \\
	&= 0.446 + {\cal O}(\lambda^6). \label{eq:23mixingNLO2}
\end{align}
Eq.\eqref{eq:23mixingNLO2} indicates $\theta_{23} < \pi/4$. It is easy to confirm that eqs.\eqref{eq:23mixingNLO} and \eqref{eq:23mixingNLO2} are consistent. The predicted value of $\theta_{23}$ is $\theta_{23} \simeq 41.9^\circ$. 

\begin{figure}[tbp]
\centering 
\hfill
\includegraphics[width=1\textwidth,origin=c]{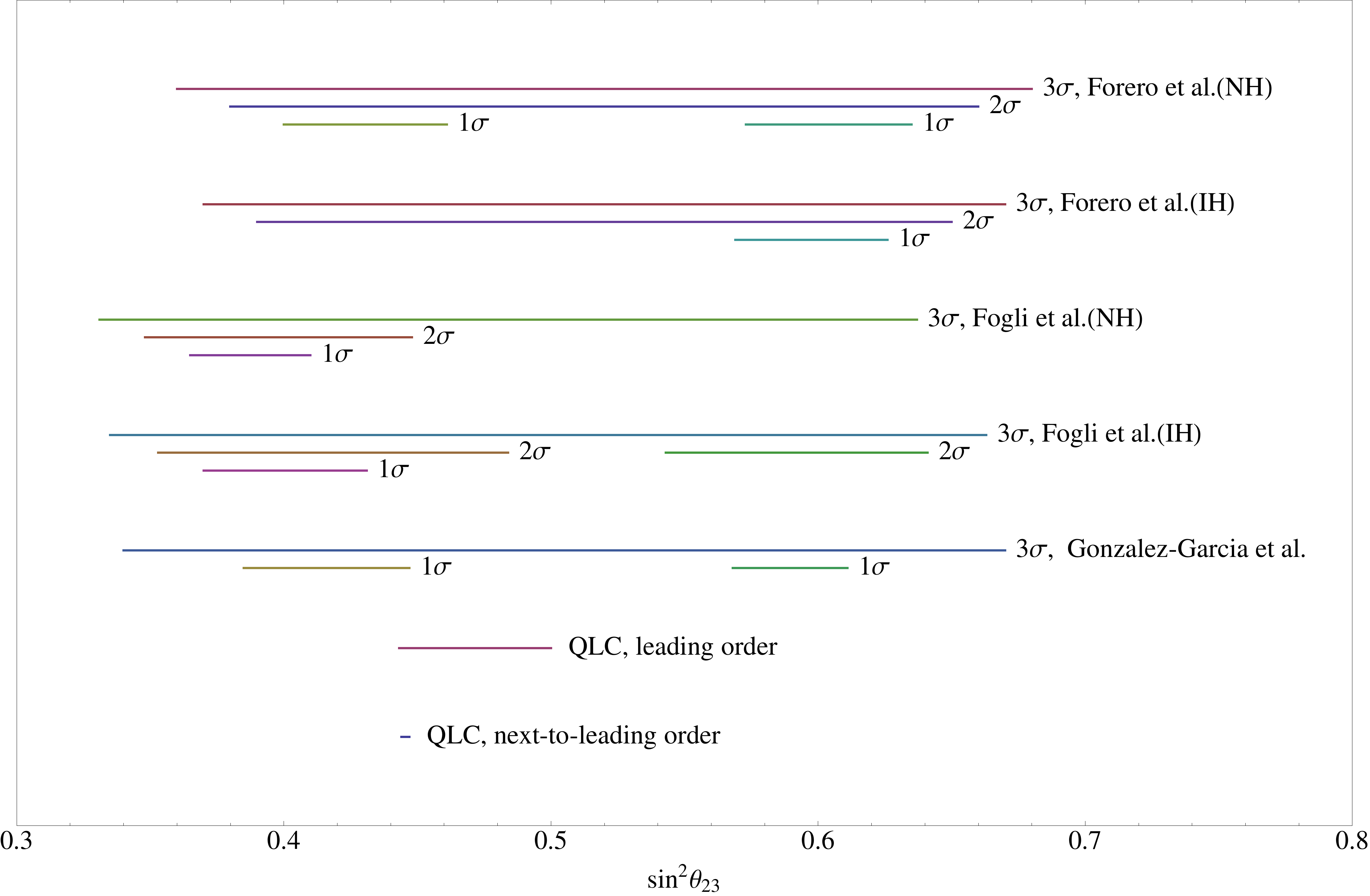}
\caption{\label{fig:23mixing} Determination of the 2-3 mixing. Shown are the results of global analysis from Forero et al.\cite{Tortola:2012te}, Fogli et al.\cite{Fogli:2012ua} and Gonzalez-Garcia et al.\cite{GonzalezGarcia:2012sz} for the normal and inverted hierarchy cases. The QLC predictions are shown at leading and next-to-leading order.}
\end{figure}

In figure~\ref{fig:23mixing}, we summarize the values of $\sin^2 \theta_{23}$. The results of global analysis of Forero et al.\cite{Tortola:2012te} and Fogli et al.\cite{Fogli:2012ua} are shown for the normal (NH) and inverted (IH) hierarchy cases. For the QLC predictions, eq.\eqref{eq:23mixingLO} is shown for $1 - 5\lambda^4 < \sin^2 2\theta_{23} < 1$ (leading order), and eq.\eqref{eq:23mixingNLO} is shown for $\sin^2 2\theta_{23} = 0.988 \pm 5\lambda^6$ (next-to-leading order), which correspond to $0.443 < \sin^2 \theta_{23} < 0.500$ and $0.444 < \sin^2 \theta_{23} < 0.447$, respectively. Figure~\ref{fig:23mixing} shows that the QLC prediction at next-to-leading order is consistent with the global fits\cite{Tortola:2012te,Fogli:2012ua,GonzalezGarcia:2012sz} for $\theta_{23} < \pi /4$ at the 2$\sigma$ level. 

\subsection{1-2 mixing $\theta_{12}$ \label{subsec:theta_12}}
It has been known that a deviation of $\theta_{12}$ from maximal mixing is large\cite{Altarelli:2004za,Mohapatra:2005wg,Mohapatra:2006gs,Strumia:2006db,GonzalezGarcia:2007ib, Hall:2013yha}. 
This can be naturally obtained in the model of eq.\eqref{eq:PMNS-CKM}: eq.\eqref{eq:12mixingLO} shows that a deviation of $\theta_{12}$ from maximal mixing is not ${\cal O}(\lambda^4)$ but the ${\cal O}(\lambda^2)$ effect, and therefore a large deviation is obtained. 

The 1-2 mixing angle is given by
\begin{align}
	\sin^2 2\theta_{12} = \frac{4 |U_{e1}|^2 |U_{e2}|^2}{\left( 1 - |U_{e3}|^2 \right)^2}. \label{eq:sin22theta_12}
\end{align}
Using eqs.\eqref{eq:Ue1^2}, \eqref{eq:Ue2^2} and \eqref{eq:1/(1-Ue32)^2}, we find that
\begin{align}
	\sin^2 2\theta_{12} 
	&= 1 - 2\lambda^2 - 4A(1 - \overline{\rho})\lambda^4 + {\cal O}(\lambda^6), \nonumber \\
	&= 1 - \underbrace{0.1016}_{\lambda^2 \ \mbox{\scriptsize term}} - \underbrace{0.0073}_{\lambda^4 \ \mbox{\scriptsize term}} + {\cal O}(\lambda^6), \nonumber \\ 
      &= 0.891 + {\cal O}(\lambda^6). \label{eq:12mixingNLO}
\end{align}
Eq.\eqref{eq:12mixingNLO} shows that the value of $\sin^2 2\theta_{12}$ is determined by three parameters, $\lambda$, $A$ and $\overline{\rho}$ to ${\cal O}(\lambda^6)$. The $\overline{\eta}$ is irrelevant up to ${\cal O}(\lambda^6)$. 

Eq.\eqref{eq:12mixingNLO} corresponds to $\theta_{12} \simeq 35.4^\circ$(an another solution of $\theta_{12} \simeq 54.6^\circ$ is excluded, because $\sin^2 \theta_{12}=|U_{e2}|^2/(1-|U_{e3}|^2) < 0.5$. The $\sin^2 \theta_{12}$ written in terms of $\lambda$, $A$ and $\overline{\rho}$ is given in the appendix). It is interesting that the value of eq.\eqref{eq:12mixingNLO} is very close to the tri-bimaximal (TBM) value of $\sin^2 2\theta_{12}=0.889$\cite{Harrison:2002er,Harrison:2002kp}.

\begin{figure}[tbp]
\centering 
\hfill
\includegraphics[width=1\textwidth,origin=c]{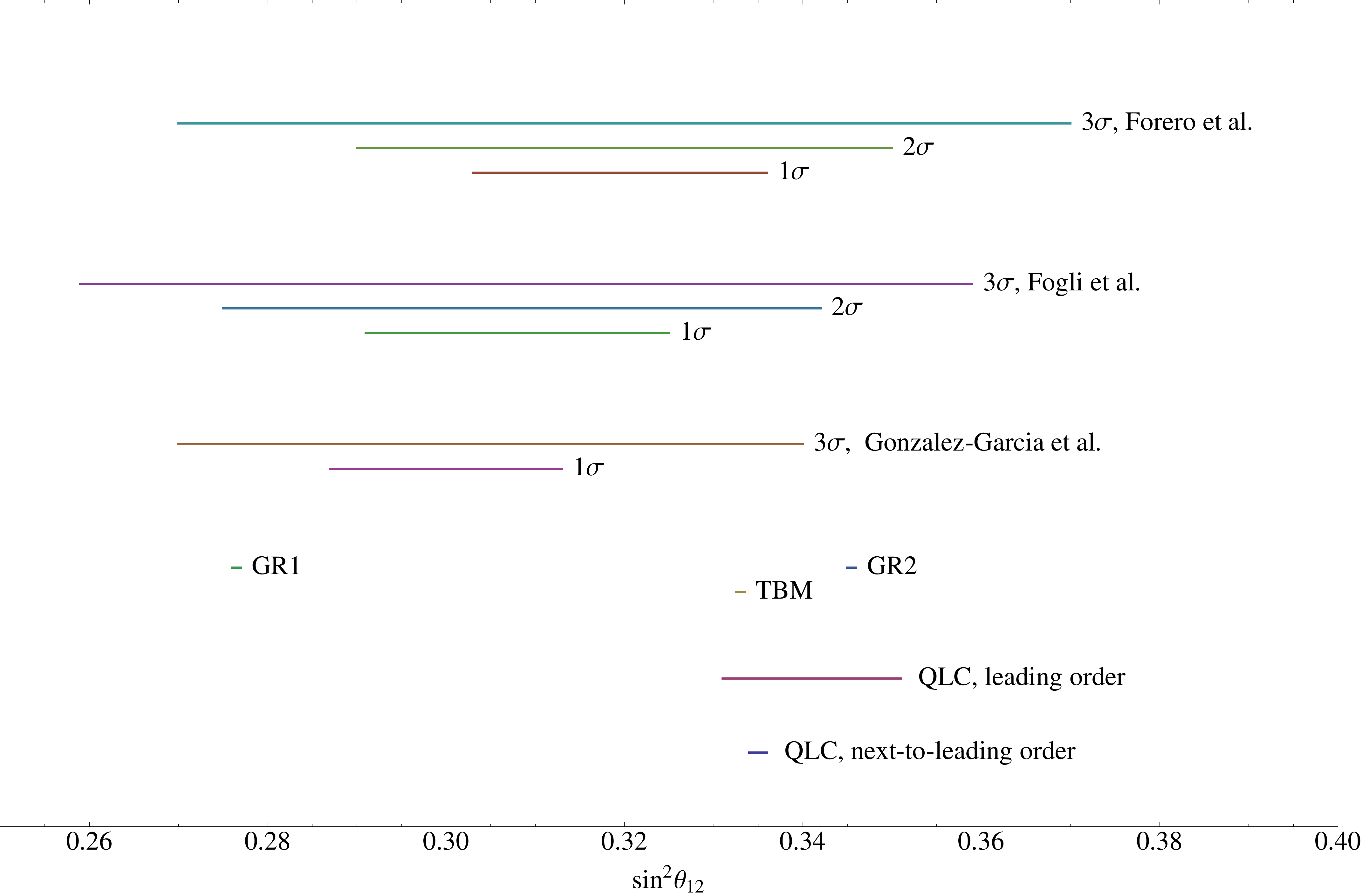}
\caption{\label{fig:12mixing} Determination of the 1-2 mixing. Shown are the results of global analysis\cite{Tortola:2012te,Fogli:2012ua,GonzalezGarcia:2012sz} and the predicted values of TBM\cite{Harrison:2002er,Harrison:2002kp}, GR1\cite{Datta:2003qg,Kajiyama:2007gx,Everett:2008et,Ding:2011cm} and GR2\cite{Rodejohann:2008ir,Adulpravitchai:2009bg,Ding:2011cm}. The QLC predictions are shown at leading and next-to-leading order.}
\end{figure}

In figure~\ref{fig:12mixing}, we summarize the values of $\sin^2 \theta_{12}$. In addition to the results of global analysis\cite{Tortola:2012te,Fogli:2012ua,GonzalezGarcia:2012sz}, the predicted TBM value of 1/3\cite{Harrison:2002er,Harrison:2002kp}, two proposed golden ratio (GR) values of 0.276(GR1\cite{Datta:2003qg,Kajiyama:2007gx,Everett:2008et,Ding:2011cm}) and 0.345(GR2\cite{Rodejohann:2008ir,Adulpravitchai:2009bg,Ding:2011cm}) are shown. For the QLC predictions, eq.\eqref{eq:12mixingLO} is shown for $\sin^2 2\theta_{12} = 0.898 \pm 5\lambda^4$ (leading order), and eq.\eqref{eq:12mixingNLO} is shown for $\sin^2 2\theta_{12} = 0.891 \pm 5 \lambda^6$ (next-to-leading order), which correspond to $0.331 < \sin^2 \theta_{12} < 0.351$ and $0.334 < \sin^2 \theta_{12} < 0.336$, respectively. Figure~\ref{fig:12mixing} shows that the QLC prediction at next-to-leading order is consistent with the global fits at the 2$\sigma$ level. 

\section{Summary and conclusions \label{sec:conclusions}}
In this paper we have presented analytical formulae of neutrino mixing angles at the next-to-leading order in the framework of eq.\eqref{eq:PMNS-CKM}. It has been shown that higher order corrections were important to explain the observed pattern of neutrino mixing. Some conclusions obtained in this paper are given below.
\begin{itemize}
	\item $\theta_{13}:$ $\sin^2 2\theta_{13}$ is determined by three parameters $\lambda$, $A$ and $\overline{\rho}$ up to ${\cal O}(\lambda^6)$. The $\overline{\eta}$ is irrelevant up to ${\cal O}(\lambda^6)$. The ${\cal O}(\lambda^4)$ corrections reduce the value of $\sin^2 2\theta_{13}$ from 0.102 to 0.092, which is consistent with five experiments and three global fits at the $1\sigma$ level. A summary is shown in figure~\ref{fig:13mixing}. 
	\item $\theta_{23}:$ the ${\cal O}(\lambda^4)$ corrections lead to a deviation of $\theta_{23}$ from maximal mixing. A deviation of $\sin^2 2\theta_{23}$ from 1 is determined only by two parameters $\lambda$ and $A$ up to ${\cal O}(\lambda^6)$. The negative sign of $\theta_{23} - \pi/4$ is predicted, and the obtained value of $\sin^2 \theta_{12}$ is 0.446, which is consistent with three global fits at the $2\sigma$ level. A summary is shown in figure~\ref{fig:23mixing}.
	\item $\theta_{12}:$ $\sin^2 2\theta_{12}$ is determined by three parameters $\lambda$, $A$ and $\overline{\rho}$ up to ${\cal O}(\lambda^6)$. The $\overline{\eta}$ is irrelevant up to ${\cal O}(\lambda^6)$. The predicted value of $\sin^2 \theta_{12}$ at next-to-leading order is 0.335, which is very close to the TBM value of 0.333. These values are consistent with the global fits at the $2\sigma$ level. A summary is shown in figure~\ref{fig:12mixing}.
\end{itemize}

In this paper we have shown that eq.\eqref{eq:PMNS-CKM} is consistent with experimental results with high precision. If it will be confirmed by  experiments that some other corrections from eq.\eqref{eq:PMNS-CKM} are very small or negligible, it may indicate that there exists an unknown theoretical mechanism behind eq.\eqref{eq:PMNS-CKM}. Therefore, it is very interesting to test eq.\eqref{eq:PMNS-CKM} by near future experiments.

\section{Appendix: The matrix elements of the PMNS matrix \label{app:PMNSelements}}
In this appendix we present the PMNS matrix elements $U_{\alpha i}$$(\alpha = e,\mu,\tau, i=1,2,3)$. We use the Wolfenstein parametrization\cite{Wolfenstein:1983yz,Beringer:1900zz} of $V_{\mbox{\scriptsize CKM}}$. 

The $V_{\mbox{\scriptsize CKM}}$ to ${\cal O}(\lambda^4)$ has been widely used in the literature, however, $V_{\mbox{\scriptsize CKM}}$ to ${\cal O}(\lambda^6)$  is necessary for calculations at next-to-leading order.
We can write $V_{\mbox{\scriptsize CKM}}$ to ${\cal O}(\lambda^6)$ in terms of the Wolfenstein parameters $\lambda$, $A$, $\overline{\rho}$ and $\overline{\eta}$, 
\begin{align}
	 V_{\mbox{\scriptsize CKM}} &= 
	\begin{pmatrix}
		V_{ud} & V_{us} & V_{ub} \\
		V_{cd} & V_{cs} & V_{cb} \\
		V_{td} & V_{ts} & V_{tb}
	\end{pmatrix},
\end{align}	
\begin{align}
	V_{ud} & = 1-\frac{\lambda^2}{2} - \frac{\lambda^4}{8} + {\cal O}(\lambda^6), \\
	V_{us} & = \lambda +{\cal O}(\lambda^6), \\
	V_{ub} & = A\lambda^3\left(1+\frac{\lambda^2}{2}\right)(\overline{\rho} - i \overline{\eta}) +{\cal O}(\lambda^6), \\
	V_{cd} & = -\lambda +A^2 \lambda^5 \left( \frac{1}{2} - \overline{\rho} - i \overline{\eta} \right) +{\cal O}(\lambda^6), \\
	V_{cs} & = 1-\frac{\lambda^2}{2} - \frac{\lambda^4}{8}\left( 1 + 4A^2 \right) +{\cal O}(\lambda^6), \\
	V_{cb} & = A\lambda^2 +{\cal O}(\lambda^6), \\
	V_{td} & = A\lambda^3(1-\overline{\rho} - i \overline{\eta}) +{\cal O}(\lambda^6), \\
	V_{ts} & = -A\lambda^2 + A\lambda^4(\frac{1}{2}-\overline{\rho} - i\overline{\eta}) +{\cal O}(\lambda^6), \\
	V_{tb} & = 1 - \frac{A^2\lambda^4}{2} +{\cal O}(\lambda^6),
\end{align}
where the Wolfenstein parameters $\lambda$, $A$, $\overline{\rho}$ and $\overline{\eta}$ are defined by 
\begin{align}
	s_{12} = \lambda, \quad s_{23} = A\lambda^2, \quad
	s_{13}e^{i\delta} = \frac{A\lambda^3(\overline{\rho} + i\overline{\eta})\sqrt{1-A^2\lambda^4}}{\sqrt{1-\lambda^2}[1-A^2\lambda^4(\overline{\rho}+i\overline{\eta})]}.
\end{align}
The quark mixing angles $s_{ij} = \sin \theta_{ij}$ and the Kobayashi-Maskawa CP phase $\delta$ are defined by the standard parametrization.
The CKM matrix written in terms of $\lambda$, $A$, $\overline{\rho}$ and $\overline{\eta}$ is unitary to all orders in $\lambda$.
The values of the Wolfenstein parameters are given by\cite{Beringer:1900zz} 
\begin{align}
	\lambda = 0.22535 \pm 0 .00065, \quad 
	A = 0.811^{+0.022}_{-0.012}, \quad
	\overline{\rho} = 0.131^{+0.026}_{-0.013}, \quad
	\overline{\eta} = 0.345^{+0.013}_{-0.014}. \label{eq:Wolfenstein_values}
\end{align}

The PMNS mixing matrix in the model of eq.\eqref{eq:PMNS-CKM} is
\begin{align}
	 U_{\mbox{\scriptsize PMNS}} = 
	\begin{pmatrix}
		U_{e1} & U_{e2} & U_{e3} \\
		U_{\mu 1} & U_{\mu 2} & U_{\mu 3} \\
		U_{\tau 1} & U_{\tau 2} & U_{\tau 3}
	\end{pmatrix}
	=
	V_{\mbox{\scriptsize CKM}}^\dagger V_{bm}, \quad
	\label{eq:PMNS-bimaximal}	
	V_{bm} =
	\begin{pmatrix}
		\frac{1}{\sqrt{2}} & \frac{1}{\sqrt{2}} & 0 \\
		-\frac{1}{2} & \frac{1}{2} & \frac{1}{\sqrt{2}} \\
		\frac{1}{2} & - \frac{1}{2} & \frac{1}{\sqrt{2}}
	\end{pmatrix}.
\end{align}
We summarize the PMNS matrix elements $U_{\alpha i} (\alpha = e, \mu, \tau, i = 1, 2, 3)$ and the squared $|U_{\alpha i}|^2$ to ${\cal O}(\lambda^6)$:
\begin{align}
	U_{e1} &= 
			\frac{1}{\sqrt{2}} 
			+ \frac{\lambda}{2} 
			- \frac{\lambda^2}{2\sqrt{2}}
			+\frac{A\lambda^3}{2}(1 - \overline{\rho} + i \overline{\eta})
		       - \frac{\lambda^4}{8\sqrt{2}}
			-\frac{A^2 \lambda^5}{2} \left(\frac{1}{2} - \overline{\rho} + i \overline{\eta} \right) 
			+ {\cal O}(\lambda^6),
			\label{eq:Ue1}\\
	U_{e2} &= 
			\frac{1}{\sqrt{2}} 
			- \frac{\lambda}{2} - \frac{\lambda^2}{2\sqrt{2}}
			-\frac{A\lambda^3}{2}(1 - \overline{\rho} + i \overline{\eta})
		       - \frac{\lambda^4}{8\sqrt{2}}
			+\frac{A^2 \lambda^5}{2} \left(\frac{1}{2} - \overline{\rho} + i \overline{\eta} \right) 
			+ {\cal O}(\lambda^6),
			\label{eq:Ue2}\\		
	U_{e3} &= 
			- \frac{\lambda}{\sqrt{2}} 
			+ \frac{A\lambda^3}{\sqrt{2}}(1-\overline{\rho}+i\overline{\eta})
		       +\frac{A^2 \lambda^5}{\sqrt{2}} \left(\frac{1}{2} - \overline{\rho} + i \overline{\eta} \right) 
			+ {\cal O}(\lambda^6),
			\label{eq:Ue3} \\	
	U_{\mu 1} &= 
			-\frac{1}{2} 
			+ \frac{\lambda}{\sqrt{2}} 
			+ \frac{\lambda^2}{2}\left(\frac{1}{2} -  A \right)
		       +\frac{\lambda^4}{2}\left\{\frac{1}{8} + \frac{A^2}{2} + A\left(\frac{1}{2} - \overline{\rho} + i \overline{\eta}\right) \right\} 
			+ {\cal O}(\lambda^6),
			\label{eq:Umu1}\\
	U_{\mu 2} &= 
			\frac{1}{2} 
			+ \frac{\lambda}{\sqrt{2}} 
			- \frac{\lambda^2}{2}\left(\frac{1}{2} -  A \right)
		 	-\frac{\lambda^4}{2}\left\{\frac{1}{8} + \frac{A^2}{2} + A\left(\frac{1}{2} - \overline{\rho} + i \overline{\eta}\right) \right\} 
			+ {\cal O}(\lambda^6),
			\label{eq:Umu2}\\
	U_{\mu 3} &= 
			\frac{1}{\sqrt{2}} 
			- \frac{\lambda^2}{\sqrt{2}}\left( \frac{1}{2} + A \right)
		       -\frac{\lambda^4}{\sqrt{2}}\left\{\frac{1}{8} + \frac{A^2}{2} - A\left( \frac{1}{2} - \overline{\rho} + i \overline{\eta} \right)\right\} 
			+ {\cal O}(\lambda^6),
			\label{eq:Umu3}\\
	U_{\tau 1} &= 
			\frac{1}{2} 
			- \frac{A\lambda^2}{2} 
			+ \frac{A\lambda^3}{\sqrt{2}} (\overline{\rho} + i\overline{\eta}) 
			- \frac{A^2 \lambda^4}{4}
		       +\frac{A\lambda^5}{2\sqrt{2}} (\overline{\rho} + i\overline{\eta} ) 
			+ {\cal O}(\lambda^6),
			\label{eq:Utau1}\\
	U_{\tau 2} &= 
			-\frac{1}{2} 
			+ \frac{A\lambda^2}{2} 
			+ \frac{A\lambda^3}{\sqrt{2}} (\overline{\rho} + i\overline{\eta}) 
			+ \frac{A^2 \lambda^4}{4}
		       +\frac{A\lambda^5}{2\sqrt{2}} (\overline{\rho} + i\overline{\eta} ) 
			+ {\cal O}(\lambda^6),
			\label{eq:Utau2}\\
	U_{\tau 3} &= 
			\frac{1}{\sqrt{2}} 
			+ \frac{A\lambda^2}{\sqrt{2}} 
			- \frac{A^2 \lambda^4}{2\sqrt{2}} 
			+ {\cal O}(\lambda^6);
			\label{eq:Utau3}
\end{align}
\begin{align}
	|U_{e1}|^2 &= 
			\frac{1}{2} 
			+ \frac{\lambda}{\sqrt{2}} 
			- \frac{\lambda^2}{4} 
			- \frac{\lambda^3}{\sqrt{2}}\left\{\frac{1}{2} - A(1-\overline{\rho})\right\}
			+\frac{A\lambda^4}{2}(1 - \overline{\rho}) \nonumber \\
			&
			- \frac{\lambda^5}{8\sqrt{2}}(1+2A)\left\{1 + 2A(1 - 2\overline{\rho})\right\}  
			+ {\cal O}(\lambda^6), 
			\label{eq:Ue1^2}\\
	|U_{e2}|^2 &= 
			\frac{1}{2} 
			- \frac{\lambda}{\sqrt{2}} 
			- \frac{\lambda^2}{4} 
			+ \frac{\lambda^3}{\sqrt{2}}\left\{\frac{1}{2} - A(1-\overline{\rho})\right\}
			+\frac{A\lambda^4}{2}(1 - \overline{\rho}) \nonumber \\
			&
			+ \frac{\lambda^5}{8\sqrt{2}}(1+2A)\left\{1 + 2A(1 - 2\overline{\rho})\right\}  
			+ {\cal O}(\lambda^6)
			\label{eq:Ue2^2}, \\
	|U_{e3}|^2 &= \frac{\lambda^2}{2} - A\lambda^4(1-\overline{\rho}) + {\cal O}(\lambda^6),	\label{eq:Ue3^2}
\end{align}
\begin{align}
	|U_{\mu 1}|^2 &= 
			\frac{1}{4} 
			- \frac{\lambda}{\sqrt{2}} 
			+ \frac{\lambda^2}{2}\left(\frac{1}{2} + A \right) 
			+\frac{\lambda^3}{\sqrt{2}}\left(\frac{1}{2} - A \right) 
			- \frac{A\lambda^4}{2}\left(1 - \overline{\rho} \right) \nonumber \\
			&
			+\frac{\lambda^5}{\sqrt{2}}\left\{\frac{1}{8} + \frac{A^2}{2} + A\left(\frac{1}{2} - \overline{\rho}\right)\right\} 
			+ {\cal O}(\lambda^6), \label{eq:Umu1^2}\\
	|U_{\mu 2}|^2 &= 
			\frac{1}{4} 
			+ \frac{\lambda}{\sqrt{2}} 
			+ \frac{\lambda^2}{2}\left(\frac{1}{2} + A \right) 
			-\frac{\lambda^3}{\sqrt{2}}\left(\frac{1}{2} - A \right) 
			- \frac{A\lambda^4}{2}\left(1 - \overline{\rho} \right) \nonumber \\
			&
			-\frac{\lambda^5}{\sqrt{2}}\left\{\frac{1}{8} + \frac{A^2}{2} + A\left(\frac{1}{2} - \overline{\rho}\right)\right\} 
			+ {\cal O}(\lambda^6), \label{eq:Umu2^2}\\
	|U_{\mu 3}|^2 &= 
			\frac{1}{2} 
			- \lambda^2\left( \frac{1}{2} + A \right)
			+ A\lambda^4\left( 1 - \overline{\rho} \right) 
			+ {\cal O}(\lambda^6), 
			\label{eq:Umu3^2}
\end{align}
\begin{align}
	|U_{\tau 1}|^2 &= 
			\frac{1}{4} 
			- \frac{A\lambda^2}{2} 
			+ \frac{A\lambda^3 \overline{\rho}}{\sqrt{2}} 
			+ \frac{A\lambda^5\overline{\rho}}{\sqrt{2}} \left( \frac{1}{2} - A \right) 
			+ {\cal O}(\lambda^6), \label{eq:Utau1^2}\\
	|U_{\tau 2}|^2 &= 
			\frac{1}{4} 
			- \frac{A\lambda^2}{2} 
			- \frac{A\lambda^3 \overline{\rho}}{\sqrt{2}} 
			- \frac{A\lambda^5\overline{\rho}}{\sqrt{2}} \left( \frac{1}{2} - A \right) 
			+ {\cal O}(\lambda^6), \label{eq:Utau2^2}\\
	|U_{\tau 3}|^2 &= 
			\frac{1}{2} 
			+ A\lambda^2 
			+ {\cal O}(\lambda^6). \label{eq:Utau3^2}
\end{align}
For convenience, we present $1/(1-|U_{e3}|^2)$, $1/(1-|U_{e3}|^2)^2$ and $\sin^2 \theta_{12}$ in terms of $\lambda$, $A$ and $\overline{\rho}$,
\begin{align}
	\frac{1}{1-|U_{e3}|^2} &= 
			1 
			+ \frac{\lambda^2}{2}
			+ \left\{ \frac{1}{4} - A (1 - \overline{\rho}) \right\}\lambda^4 
			+ {\cal O}(\lambda^6), \label{eq:1/1-Ue32} \\
	\frac{1}{\left(1-|U_{e3}|^2\right)^2} &= 
			1 
			+ \lambda^2
			+ \left\{ \frac{3}{4} - 2A (1 - \overline{\rho}) \right\}\lambda^4 
			+ {\cal O}(\lambda^6), \label{eq:1/(1-Ue32)^2} \\
	\sin^2 \theta_{12} &= \frac{1}{2} - \frac{\lambda}{\sqrt{2}} - \frac{A}{\sqrt{2}}(1 - \overline{\rho})\lambda^3
	+ \frac{\lambda^5}{\sqrt{2}}\left\{\frac{1}{8} + A(1-\overline{\rho}) + A^2 \left(\frac{1}{2} - \overline{\rho}\right)\right\} 
	+ {\cal O}(\lambda^6).
\end{align}

\bibliography{references}

\end{document}